\documentclass[pre,aps,twocolumn
,showpacs,10pt]{revtex4-1}

\usepackage[pdftex]
{graphicx}
\usepackage{float}
\usepackage[caption = false]{subfig}
\usepackage{amssymb,amsfonts,amsmath}
\usepackage{xcolor}

\begin{document}

\title{Isotropic-nematic transition of self-propelled rods in three dimensions}

\author{M.C.~Bott}
\affiliation{Soft Matter Theory, University of Fribourg, CH-1700 Fribourg, Switzerland}
\email{matthiaschristian.bott@unifr.ch}

\author{F.\ Winterhalter}
\affiliation{Institut f\"ur Theoretische Physik, Universit\"at Erlangen-N\"urnberg, 91058 Erlangen, Germany}

\author{M.\ Marechal}
\affiliation{Institut f\"ur Theoretische Physik, Universit\"at Erlangen-N\"urnberg, 91058 Erlangen, Germany}

\author{A.\ Sharma}
\affiliation{Leibniz-Institut f\"ur Polymerforschung Dresden, 01069 Dresden, Germany}

\author{J.M.~Brader}
\affiliation{Soft Matter Theory, University of Fribourg, CH-1700 Fribourg, Switzerland}

\author{R.\ Wittmann}
\affiliation{Soft Matter Theory, University of Fribourg, CH-1700 Fribourg, Switzerland}

\begin{abstract}
Using overdamped Brownian dynamics simulations we investigate the isotropic--nematic 
(IN) transition of self-propelled rods in three spatial dimensions. 
For two well-known model systems (Gay-Berne potential and hard spherocylinders) we 
find that turning on activity moves to higher densities the phase boundary separating  
an isotropic phase from a (nonpolar) nematic phase. 
This active IN phase boundary is distinct from the 
 boundary between isotropic and polar-cluster states previously 
reported in two-dimensional simulation studies and, unlike the latter, is not sensitive 
to the system size. 
We thus identify a generic feature of anisotropic active particles in three dimensions. 
\end{abstract}

\maketitle

\section{Introduction}

Collective nonequilibrium behavior in suspensions of active Brownian particles (ABPs) is the subject of much 
current research interest \cite{Romanczuk2012}. 
Not only do these systems exhibit novel dynamics and phase behavior, they are 
also relevant for understanding self-organization phenomena in nature. 
Much of the interest in ABPs has been driven by the introduction of 
new experimental model systems,
such as catalytic Janus particles \cite{Palacci2010,Erbe2008,Howse2007}, light 
activated colloids~\cite{Palacci2013} and colloids with artificial flagella \cite{Dreyfus2005}. 
Additionally, studies of minimal spherical active models have triggered a whole new branch of fundamental research 
in nonequilibrium statistical mechanics.
The striking similarities to an equilibrium system have been exploited by
developing a Cahn-Hilliard-like mechanism \cite{Speck2014} to describe the early-stage dynamics of motility-induced phase separation \cite{Tailleur2008,Buttinoni2013,Cates2015},
identifying an effective equilibrium regime \cite{Maggi2015,Farage2015,Fodor2016,Wittmann2017},
defining effective interaction potentials~\cite{Cates2015,Schwarz-Linek2012,Wittmann2016,Wittmann2017}
or employing linear-response theory \cite{Sharma2016}.
More fundamentally, a better understanding of active 
pressure \cite{Takatori2014,Solon2015} or chemical potential \cite{Paliwal2018} is required
to provide a solid framework for active thermodynamics \cite{Speck2016,Solon2018}.
 Recently, also the question of how activity influences the well-studied phase transitions in a passive system of soft disks has been addressed in detail \cite{kapfer}.

While spherical ABPs are ideal for exploring basic concepts, 
suspensions of \textit{anisotropic} ABPs are perhaps more relevant, 
as these better represent the generic type of particles encountered in nature 
\cite{Elgeti2015,Marchetti2013}. 
Self-propelled rods (SPRs), the anisotropic analog of ABPs, for which the self propulsion along 
the long axis of the particle breaks the up-down symmetry, exhibit a rich dynamical phase behavior at high (infinite)
activity \cite{Wensink2012PNAS,Wensink2012,Abkenar2013} in two dimensions (2D). 
Simulations of large 2D systems (with rotational diffusion) \cite{Yang2010,Abkenar2013,Weitz2015} reveal that
at densities below the passive isotropic--nematic (IN) transition, the initially isotropic state 
begins to destabilize due to the emergence of moving polar clusters, which grow in size upon increasing 
activity but do not form a global phase \cite{Weitz2015}. At higher densities, a laning phase is found,
which does have nematic order on the range of the simulation domain, but is not homogeneous~\cite{Wensink2012PNAS,Wensink2012,Abkenar2013}.
Experiments on a fluidized monolayer of rods 
have identified giant number fluctuations in such states \cite{Narayan2007}. 
 For experiments on very long (and thus non-Brownian) bacteria in quasi-2D, a nematic phase with long-range order was reported~\cite{Nishiguchi2017}.
 
The (enhanced) nematic ordering of a biologically inspired 2D nematic model has been studied in 
simulation \cite{Kraikivski2006}. In extensions of the Vicsek model to incorporate local nematic ordering (rather than the polar ordering of the original Vicsek model), the region of stability of homogeneous active nematic phase (with giant
number fluctuations) in 2D is determined~\cite{Ginelli2010,Ngo2014}.
It is unsure whether this phase has 
long-range or quasi-long-range order in these agent-based simulations; it even seems to depend on the details of the model (compare Refs.~\onlinecite{Ginelli2010} and~\onlinecite{Ngo2014}).
 Returning to the SPRs interaction model, the lack of observations of a homogeneous nematic phase in previous simulations (in 2D) means that the IN phase boundary 
for overdamped SPRs has not been addressed explicitly. 
 To avoid confusion of terminology we emphasize that in the literature
`active nematics' usually address anisotropic particles driven randomly 
back and forth along their axis~\cite{Marchetti2013}. In this sense the term `nematic' 
refers to the particle symmetry. 
The present work concerns SPRs and the terms `polar' and 
`nematic' will be reserved to describe collective states.

The theoretical understanding of the phase behavior of SPRs is a difficult problem. 
According to an early mean-field approach \cite{Baskaran2008a}, the density at the IN phase transition is insensitive to activity. 
In contrast, more general collision-based models \cite{Baskaran2008,Baskaran2010} predict that the transition density decreases with increasing activity. 
For overdamped (Langevin) dynamics, the current numerical evidence 
suggesting that activity might stabilize nematic order of SPRs 
only arises from the observation that, 
as the density increases, the destabilization of the isotropic phase with respect to polar fluctuations 
occurs at lower activities \cite{Yang2010,Weitz2015}. 
In general, the existence of a (nonpolar) nematic phase and its phase boundary 
remains an open problem.

\begin{figure}[t!]
\includegraphics[width=0.48\textwidth]{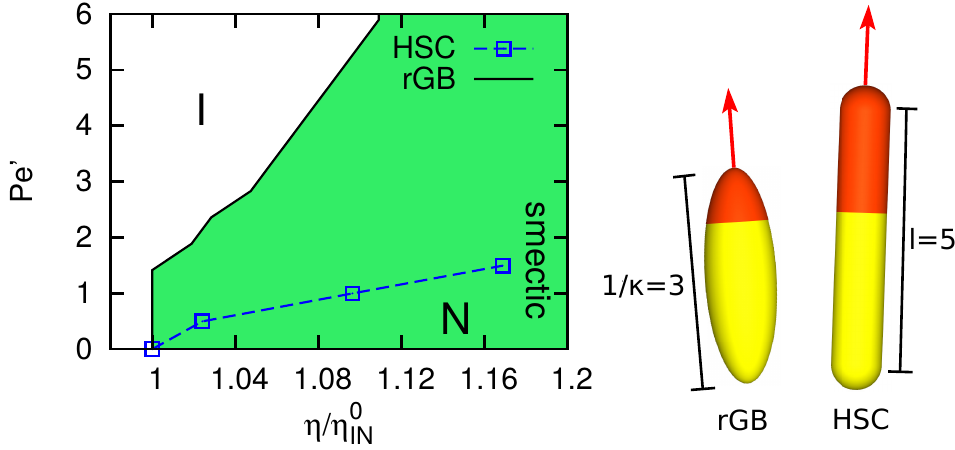}
\caption{
 Isotropic--nematic (IN) transition of self-propelled rods (SPRs).
The solid line and the filled region corresponds to soft-repulsive Gay-Berne (rGB) ellipsoids
and the dashed line and symbols to hard spherocylinders (HSC).
The particles, sketched on the right indicating their (approximate) length-to-width ratio,
are propelled with constant velocity $v_0$ in the direction of their orientation $\mathbf{\hat{u}}_i$ (red arrows).
In both systems, the state points are given by the
 modified P\'eclet number $\text{Pe}'=v_0/\sqrt{6 D_\text{T}
D_\text{R}}$ (swim speed
rescaled by a particle-geometry dependent factor) and
the packing fraction $\eta$, relative to the IN transition packing fraction $\eta_\mathrm{IN}^0$ in equilibrium. 
 Beyond a certain threshold $\eta$ (as indicated by the label ``smectic''), we find
smectic clusters for HSC (more work is required to determine whether these are a separate phase).
\label{fig:HSC_GB_comparison}}
\end{figure}

In this paper we address the activity dependence of the IN transition of SPRs close to equilibrium using 
overdamped Brownian dynamics simulations. We have chosen to mostly work in three dimensions (3D), because \emph{(i)} we expected a greater stability of the homogeneous nematic phase
in 3D than in 2D (as for the phase-separated state of ABPs~\cite{Stenhammar2014}) \emph{(ii)} giant number fluctuations are predicted to be reduced in 3D compared to 2D for active nematics~\cite{Ramaswamy2003}
(it should be noted that the validity of the linearized theory from which these predictions stem is debated for 2D systems~\cite{Ginelli2010,Ngo2014,Shankar2018})
and \emph{(iii)} the equilibrium nematic phase has long-ranged order in 3D (as opposed to quasi-long-range order in 2D), 
which makes it easier to describe using particle-resolved computer simulations and other theoretical approaches---leading to less finite-size effects 
in the former---than in 2D.
We will further consider systems of lower aspect ratio, well away from the Onsager limit. 

Although we will focus on the repulsive Gay-Berne (rGB) model,
which is a classic model of thermotropic liquid crystals \cite{Gay1981,Adams1987,Luckhurst1990}, 
we have also performed simulations of lyotropic hard 
spherocylinders (HSC) \cite{Bolhuis1997,Wittmann2016a}
to ensure that our findings are not model-specific.
For both of the considered models we observe a shift of the IN transition towards higher 
densities as the system is driven out of equilibrium by turning on the activity, see Fig.~\ref{fig:HSC_GB_comparison}. 
The IN transition line is independent of system size, in contrast to the 
transition between isotropic and polar-cluster states found at higher activities. 
A preliminary survey of simulations of the rGB model in 2D (see the supplementary information (SI)~\cite{supinfo_active_ellipsoids})
indicates that nematic order is more rapidly disrupted by turning on activity than in 3D, 
but finite size effects preclude a definite statement.
 This might be the reason why such an active IN transition has not been reported earlier.
The remainder of this paper is arranged as follows.
 In Sec. \ref{sec:model} we present the two different models we considered
and describe the simulation method employed for each of them.
In Sec. \ref{sec:results} we present the simulation results of the two models and identify the nematic phase at finite activity.
Finally, in Sec. \ref{sec:conclusion} we discuss our findings and provide an outlook.

\section{Simulation methods}
\label{sec:model}

\subsection{Gay-Berne model}

The regular Gay-Berne potential~\cite{Gay1981,Adams1987,Luckhurst1990} between a pair of particles 
is of the form of a Lennard-Jones potential, whose depth and range depend on the inter-particle separation and the particle orientations:
\begin{align}
 \label{eq:gay_berne_potential}
 \phi_{\rm gb}(\mathbf{\hat{u}}_1, \mathbf{\hat{u}}_2, \mathbf{r}) &= \nonumber \\
 &4\epsilon(\mathbf{\hat{u}}_1, \mathbf{\hat{u}}_2, \mathbf{\hat{r}})
 \Biggl\{\biggl( \frac{\sigma_0}{r - \sigma(\mathbf{\hat{u}}_1, \mathbf{\hat{u}}_2, \mathbf{\hat{r}}) + \sigma_0}\biggr)^{12} \nonumber \\ 
 &- \biggl( \frac{\sigma_0}{r - \sigma(\mathbf{\hat{u}}_1, \mathbf{\hat{u}}_2, \mathbf{\hat{r}}) + \sigma_0}\biggr)^{6} \Biggr\}.
\end{align}
Here the unit vectors $\mathbf{\hat{u}}_1$ and $\mathbf{\hat{u}}_2$ specify the orientation of the interacting particles $1$ and $2$ and $\mathbf{r}$, $r$, $\mathbf{\hat{r}}$ their center to center-vector, -distance and -direction.
The attraction depth $\epsilon(\mathbf{\hat{u}}_1, \mathbf{\hat{u}}_2, \mathbf{\hat{r}})$ and the range $\sigma(\mathbf{\hat{u}}_1, \mathbf{\hat{u}}_2, \mathbf{\hat{r}})$ of the particle interaction are dependent on the orientation. 
The legthscale is $\sigma_0$, which in our case is the width of the particle (see also appendix \ref{app:AppendixA} for detailed description).

Here we consider a purely soft-repulsive WCA-like version~\cite{Rull1995}, the rGB model
\begin{align}\label{eq:wca_gay_berne_potential}
 \phi_{\rm rgb}&(\mathbf{\hat{u}}_1,\mathbf{\hat{u}}_2,\mathbf{r}) \nonumber \\
 &=\begin{cases}
    \phi_{\rm gb}(\mathbf{\hat{u}}_1,\mathbf{\hat{u}}_2,\mathbf{r}) + \epsilon(\mathbf{\hat{u}}_1,\mathbf{\hat{u}}_2,\mathbf{r}) \;\; r\!<\!r_{\rm min}\\
    0 \qquad \qquad \qquad \qquad \qquad \qquad r \ge r_{\rm min},
    \end{cases}
\end{align}
obtained by shifting 
and truncating the Gay-Berne potential.
We study particles with a length-to-width ratio which is roughly given by $\kappa^{-1}=3$ (see the sketch in Fig.~\ref{fig:HSC_GB_comparison}).
The packing fraction is defined as $\eta = N V_E/V$, with $V$ being the volume of the simulation box and $V_E = 4\pi\sigma_0^3/(3\kappa)$ the volume of the ellipsoidal particle.
 
The position and orientation vectors evolve in time according to the coupled Langevin equations in the overdamped limit,
\begin{align}
\label{eq:equation_of_motion}
 \frac{\mathrm{d}}{\mathrm{d}t}\mathbf{{r}}_i(t) =&\; \gamma^{-1}\,\mathbf{F}_i(t)  + v_0 \mathbf{\hat{u}}_i(t) + \boldsymbol{\xi}_i(t), \\
\label{eq:equation_of_motion2}
 \frac{\mathrm{d}}{\mathrm{d}t}\mathbf{\hat{u}}_i(t) =&\; \alpha^{-1} \bigl(\mathbf{T}_i(t) \times \mathbf{\hat{u}}_i \bigr) + \bigl(\boldsymbol{\eta}_i(t) \times \mathbf{\hat{u}}_i(t)\bigr).
\end{align}
where $\gamma\!=\!k_\text{B} T/D_\text{T}$ and $\alpha\!=\!k_\text{B} T/D_\text{R}$ are friction coefficients determined by the 
translational and rotational diffusion coefficients $D_\text{T}$ and $D_\text{R}$, which we define as $D_\text{T}/\sigma_0^2 = 3 D_\text{R}$.
The force and torque acting on particle $i$ are related to the total potential energy, which we take to be 
a sum of pair potentials, $U_N=\sum_{i<j}\phi(\mathbf{\hat{u}}_i, \mathbf{\hat{u}}_j, \mathbf{r}_{ij})$, 
according to 
$\mathbf{F}_i(t) = -\nabla_i U_N$ and $\mathbf{T}_i(t) = - \mathbf{\hat{u}}_i \times  \partial U_N/\partial \mathbf{\hat{u}}_i$. 
The stochastic vectors $\boldsymbol{\xi}_i(t)$ and $\boldsymbol{\eta}_i(t)$ are Gaussian distributed with 
zero mean and have the time correlations 
$\langle \boldsymbol{\xi}_i(t)\otimes \boldsymbol{\xi}_j(t') \rangle = 2 D_\text{T} \mathbf{1} \delta_{ij} \delta(t-t')$
and
$\langle \boldsymbol{\eta}_i(t) \otimes \boldsymbol{\eta}_j(t') \rangle = 2 D_\text{R} \mathbf{1} \delta_{ij} \delta(t-t')$, where $\otimes$ refers to the dyadic product of two vectors. 
The active component of the dynamics enters via the second term in \eqref{eq:equation_of_motion}, where $v_0$ is a constant self-propulsion velocity.
We define the dimensionless time $t^*= t D_\text{T}/\sigma_0^2$, and activity as $v_0^* = v_0 \sigma_0 /D_\text{T}$.

\subsection{Hard spherocylinders (HSC)}

We also consider HSC interacting via a nearly hard-core 
potential with aspect ratio $l=5$ (well-studied in equilibrium) (see also the sketches in Fig.~\ref{fig:HSC_GB_comparison}).
The aspect ratio $l=L/\sigma_0$ of a HSC is given by the ratio of the cylinder length $L$ and the diameter $\sigma_0$ of the capping hemispheres.
Its volume is thus given by $V_\text{S}=\pi(l/4+1/6)\,\sigma_0^3$ and the packing fraction follows as $\eta = N V_\text{S}/V$.
For the system of active HSC, 
we use an existing simulation framework, the \textsf{pe} part of waLBerla \cite{Godenschwager2013}. The \textsf{pe} part is a massively parallel framework for molecular dynamics (MD) 
and a similar technique, the discrete element method; we do not use the Lattice Boltzmann technique (for which the waLBerla framework is better known) in this work.
We implemented a friction and noise term in this MD framework (while keeping the inertia term), which means that the Langevin equation we are using is not fully overdamped:
\begin{align}
	\frac{\mathrm{d}\mathbf{p}_i}{\mathrm{d}t}&= -{\Xi}_i \cdot \mathbf{v}_i +\mathbf{F}_{\mathrm S,i} +\mathbf{F}_{\mathrm R,i}
\label{eq:momentumLangevin}
\\
\frac{\mathrm{d}\mathbf{L}_i}{\mathrm{d}t}&= -\gamma_r \boldsymbol{\omega}_i +\mathbf{T}_{\mathrm S,i} +\mathbf{T}_{\mathrm R,i}\;,
\label{eq:angular_momentumLangevin}
\end{align}
where $\mathbf{p}_i=m\mathbf{v}_i$ and $\mathbf{L}_i={I}_i \cdot \boldsymbol{\omega}_i$ are the momentum and angular momentum of a particle with mass $m$ and inertia tensor ${I}_i$. 
Here $\mathbf{v}_i$ and $\boldsymbol{\omega}_i$ denote the translational and angular velocity, respectively.
The first term on the right hand side of Eq.~{(\ref{eq:momentumLangevin})} and (\ref{eq:angular_momentumLangevin}) accounts for friction due to the viscous dissipation. The translational friction tensor ${\Xi}_i$ depends on the 
translational friction coefficients $\gamma_{\|}$ and $\gamma_{\bot}$ for motion parallel and perpendicular to the symmetry axis $\mathbf{\hat{u}}_i$ of particle $i$: 
\begin{equation}
{{\Xi}}_i=\gamma_{\|}\mathbf{\hat{u}}_i\otimes\mathbf{\hat{u}}_i+\gamma_{\bot}(\boldsymbol{\mathbb{I}}-\mathbf{\hat{u}}_i\otimes\mathbf{\hat{u}}_i)\;,\label{eq:transFricTensor}
\end{equation}
(where $\mathbb I$ denotes the identity matrix). For reasons of symmetry, only angular velocities $\boldsymbol{\omega}_i$ perpendicular to the symmetry axis $\mathbf{\hat{u}}_i$ of the HSC are considered, therefore,
the rotational friction coefficient $\gamma_r$ for rotation of the particle axis suffices to describe the viscous torque. 

The subscript $\mathrm S$ in Eq.~{(\ref{eq:momentumLangevin})} and (\ref{eq:angular_momentumLangevin}) indicates the \textit{systematic} contributions to the force $\mathbf{F}_i$ and torque $\mathbf{T}_i$, respectively.
The first contribution to the systematic force ${\mathbf{F}}_{\mathrm S,i}$ are the particle interactions:
The particles interact only when they intersect. Overlaps are resolved by applying a fully elastic linear spring force model to any contact points. The restitution force $F_{\mathrm{rest}}$ in the direction of the contact
normal $\mathbf{n}$ acting at the contact point is given by
$
F_{\mathrm{rest}} = k\, \delta
$
with $k$ the stiffness of the potential and $\delta$ the penetration depth. We set the stiffness $k$ to a high value: $\beta k\sigma_{\text{HSC}}^2= 4\cdot 10^4$,
such that more than 99\% of the collisions at the higher densities have a penetration $\delta/\sigma_{\text{HSC}}<0.02$. 
The second contribution to $\mathbf{F}_{\mathrm S,i}$ models the self-propulsion: $\mathbf{F}_{\text{SP},i}= \gamma_\| v_0 \mathbf{\hat{u}}_i$.

The random contributions $\mathbf{F}_{\mathrm R,i}$ and $\mathbf{T}_{\mathrm R,i}$, originating from collisions with solvent molecules as mentioned above, have a Gaussian probability distribution. 
The corresponding correlation functions are related to the viscous friction according to the fluctuation-dissipation theorem for particles $i$ and $j$:
\newcommand{\kB}{k_{\mathrm{B}}}
\begin{align}
 \langle\mathbf{F}_{\mathrm R,i}(t)\rangle&=\langle\mathbf{T}_{\mathrm R,i}(t)\rangle=\mathbf{0}\cr
 \langle\mathbf{F}_{\mathrm R,i}(t)\otimes\mathbf{F}_{\mathrm R,j}(t')\rangle&=2 \kB T {{\Xi}}_i\,\delta_{ij} \delta(t-t')\nonumber\\
 \!\!\!\!\!\!\!\!\!\!\!\!\langle\mathbf{T}_{\mathrm R,i}(t)\otimes\mathbf{T}_{\mathrm R,j}(t')\rangle&=2 \kB T \gamma_r(\boldsymbol{\mathbb{I}}-\mathbf{\hat{u}}_i\otimes\mathbf{\hat{u}}_i)\,\delta_{ij}\delta(t-t')\!\!\!\!\!\!\!\!\!\!\!\!\nonumber\\\!\!\!\!\!\!
 \langle\mathbf{F}_{\mathrm R,i}(t)\otimes\mathbf{T}_{\mathrm R,j}(t')\rangle&={0}\!\!\!\!\!\!\!\!\!\!\!\!\!\!\!
\end{align}
with Boltzmann's constant $\kB$ and the temperature $T$. $\delta(t-t')$ represents the Dirac delta distribution, which in the case of discrete time steps of size $\mathrm{d}t$ is replaced by $\delta_{tt'}/\mathrm{d}t$. 
Since the angular velocity $\mathbf{\omega}_i$ is kept perpendicular to the symmetry axis of the HSC, only random torques $\mathbf{T}_{\mathrm R,i}$ normal to this axis are applied.

We define the mass $m$ and the moment of inertia ${I}_i$ of the particle such that the relaxation time 
$ \tau_I = \frac{m}{\bar{\gamma}}$ with $\frac{1}{\bar{\gamma}} = \frac{1}{\gamma_{\|}} + \frac{2}{\gamma_{\bot}}$
for the (linear) momentum is 100 times smaller 
than the Brownian time scale $\sigma_0^2/D_\text{T}$
and effects of the inertia are thus expected to be small. For the HSC system, the translational diffusion constant is given by $D_{\mathrm T}=k_{\mathrm B}
T/\bar\gamma$.

\begin{figure}
\begin{center}
\includegraphics[width=0.48\textwidth]{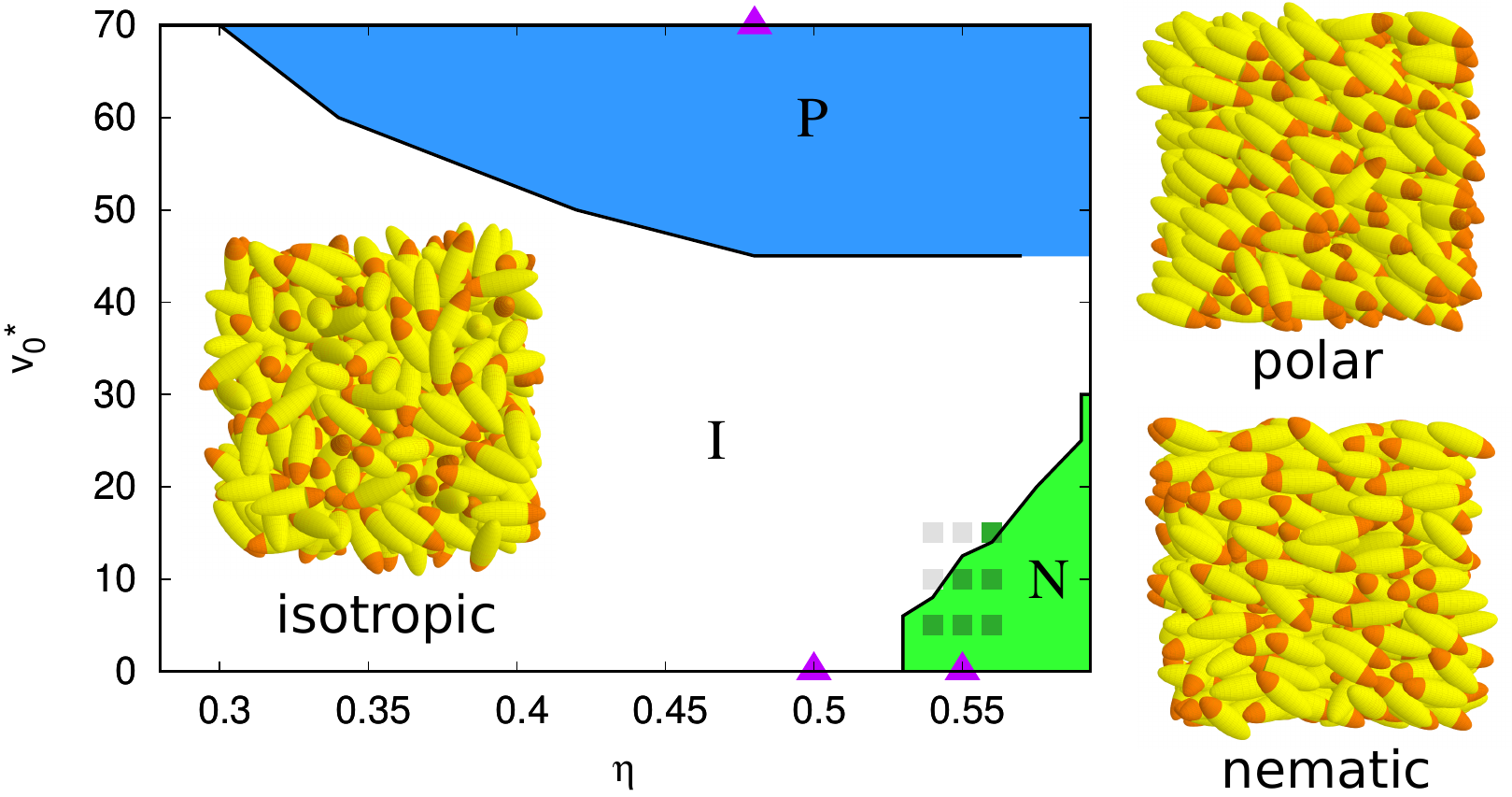}
\caption{Phase diagram of the 3D rGB model in the packing 
fraction--activity plane for $N=500$ particles.  
The stability of the isotropic (I), nematic (N) and local polar (P) states is determined by analyzing global order 
parameters.
Typical snapshots are shown for three distinct state points, as indicated by triangles. 
The IN boundary (see Fig.~\ref{fig:HSC_GB_comparison} for a closeup) is insensitive to changes in the system size and 
separates true nonequilibrium phases. 
In contrast, the boundary separating I and P states is system-size dependent;
the region P merely indicates where system-spanning polar clusters occur. 
The differently shaded squares near the IN phase boundary at $v_0^*=5,10,15$ denote the stable phase found in the larger system with $N=1000$ (see also the SI~\cite{supinfo_active_ellipsoids}).
}
\label{fig:phasediagram}
\end{center}
\end{figure}

\subsection{Parameters}
\label{sec:order}

We analyze the orientational behavior of the system by measuring the time averages $S$ and $P$ of the nematic order 
parameter $S(t)$ \cite{Eppenga1984} and the polar order parameter $P(t)$, respectively, defined at each instant of time $t$ as
\begin{equation}
 \!S(t) = \frac{1}{N} \sum_{i=1}^N \frac{3 (\mathbf{\hat{u}}_i \cdot \mathbf{\hat{n}})^2 - 1}{2}\,,\ \, P(t) = \frac{1}{N}\Big| \sum_{i=1}^N \mathbf{\hat{u}}_i \cdot \mathbf{\hat{n}} \,\Big|\,,\!\!\!\!\!\!
 \label{eq:SP}
\end{equation}
where $\mathbf{\hat{u}}_i$ is the instantaneous orientation vector of particle $i$ and $\mathbf{\hat{n}}$ 
is the nematic director, see appendix \ref{sec:app_order_pars}.
Both quantities take values between $0$ and $1$, with the extreme values indicating full disorder or perfect order, respectively.  
To distinguish between the different states we choose the 
threshold values $S_\text{t}\!=\!0.35$ for the onset of nematic and $P_\text{t}\!=\!0.55$ for polar order.
For the rGB ellipsoids, different choices for these thresholds would lead to a slight shift of the 
transition lines in the phase diagram, but not affect our main conclusions. For the HSC, the IN 
transition is more strongly discontinuous leading to a large jump in the order parameter, so the location of the IN
transition in the phase diagram is not affected by small changes of the threshold $S_\text{t}$. In the HSC system,
we never observe global polar order (due to the system size and the relatively low activities), so the value of $P_\text{t}$ is irrelevant.

In the rGB model, we simulate $N\!=\!500$ particles for various activities and densities.
The size of the simulation box is determined by the packing fraction which ranges from $\eta = 0.1$ to $\eta = 0.59$ in our computations.
This corresponds to a side length of the simulation box ranging between $b \approx 19.9\, \sigma_0$ to $b \approx 11.0\, \sigma_0$.
Our simulations yield a passive IN transition at $\eta = 0.53$, which is in agreement with 
previous studies \cite{Rull1995}.
To rule out finite-size effects, we repeated some simulation runs for $N\!=\!1000$ particles.
To show that the active IN transition is not specific to the rGB model we compare the results to larger-scale simulations of the active HSC model using $N\simeq 40000$ particles.
The IN transition in the passive HSC system lies at $\eta=0.415$, 
in good agreement with previous work~\cite{McGrother1996}.
For a finite activity we have also performed some simulations for $N=20000$ particles and observed no significant changes in the results compared to $N=40000$.

To make a proper comparison of both systems we define a dimensionless swimming speed that
does not contain arbitrary length- and time-scales. 
We thus consider the (square root of) the
active part of the single-particle diffusivity relative to the passive part~\cite{Hagen2011} and define the dimensionless
\begin{equation}
\mathrm{Pe}'=\sqrt{\frac{D_\mathrm{a}}{D_\mathrm{p}}} = \frac{v_0}{\sqrt{6 D_\mathrm{T} D_\mathrm{R}}}=\frac{v_0^*\sqrt{D_\mathrm{T}}}{\sigma_0\sqrt{6  D_\mathrm{R}}}\,.
 \end{equation}
Furthermore, we rescale the density by its value at the equilibrium IN transition for each system.

\section{Numerical results}
\label{sec:results}

For the rGB model, we mapped out an exemplary full finite-size phase diagram,
shown in Fig.~\ref{fig:phasediagram}, which reveals three distinct states in the density-activity plane: 
isotropic, nematic and polar. 
We characterize the polar state, in which the majority of particles are driven in the same direction, by $S\!>\!S_\text{t}$ and $P\!>\!P_\text{t}$.
Its occurrence here is a known artifact of a finite system \cite{Yang2010,Abkenar2013,Weitz2015}, 
which we detail below. The purpose of showing it here is to indicate the onset of large polar cluster formation (a `large' cluster contains a few hundred particles).
Outside of this region we can expect that the simulation results for the rGB model indicating an isotropic phase are trustworthy.
The large finite-size effects in this polar state are not to be confused with the distinct polar fluctuations observed at relatively low activity near the IN phase boundary, which arise due to a combination of
the finite-size effects and the enhanced tendency of the rods to align parallel within the active nematic phase,
indicate a true phase transition.

Our main result is that we observe a nonequilibrium nematic phase with $S\!>\!S_\text{t}$ but $P\!<\!P_\text{t}$,
whose boundary bends to the right with increasing $v_0$, suggesting that introducing a moderate amount of activity can suppress 
orientational ordering. 
For both the rGB and HSC systems the phenomenology depicted in Fig.~\ref{fig:HSC_GB_comparison} is consistent; the 
IN transition line moves to higher densities as the activity is increased. 
We explicitly verified that the location of this active IN transition in each model is independent of the system size,
as indicated in Fig.~\ref{fig:phasediagram}.

\begin{figure}[t!]
\includegraphics[width=0.44\textwidth]{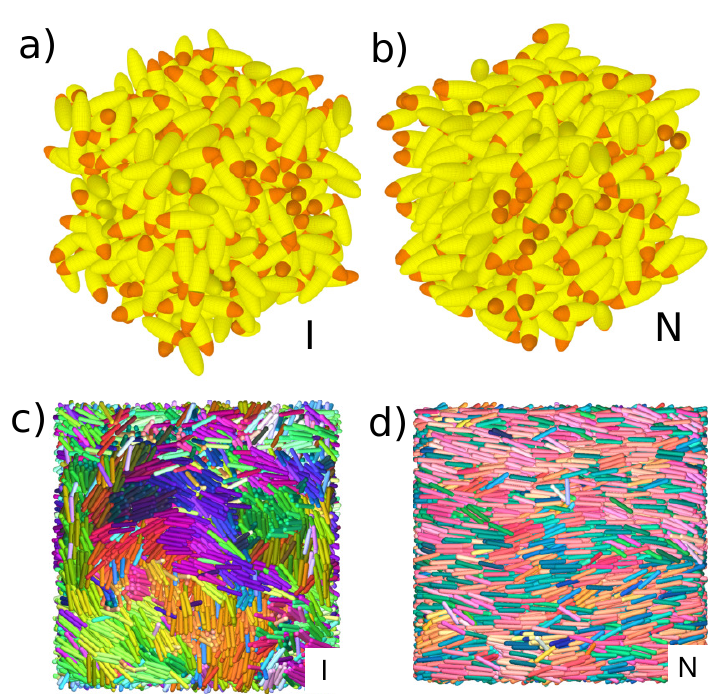}
\caption{Simulation snapshots of the isotropic (I) and nematic (N) phases for the active rGB system with $v_0^*=10$ and (a) $\eta=0.5$ and (b) $\eta=0.55$ 
and for the active HSC system (side view) at packing fractions just (c) below and (d) above the transition at $\text{Pe}' =1.5$ (see Fig.\ \ref{fig:HSC_GB_comparison}).
The color scheme serves to distinguish particles with different orientations.
}
\label{fig:simulation_snapshots}
\end{figure}

Despite the similarity of the rescaled rGB and HSC phase boundaries in Fig.~\ref{fig:HSC_GB_comparison} there are 
quantitative differences presumably related to both the interparticle interactions and the aspect ratio. 
Most notably, we observe in Fig.~\ref{fig:simulation_snapshots} a different microstructure and
the rGB system begins to exhibit local polar order as the IN phase boundary is approached.
 In addition, the equilibrium phase diagram differs in the two models in that a crystal is found for ellipsoids at higher densities, while the 
phase diagram for HSC with this aspect ratio features also
a smectic phase (that is, a phase with fluid-like layers in which the particles are nematically ordered with the director normal to the layers).
We see remnants of the latter in the active HSC system in the form of smectic clusters (not shown). These clusters usually span the system. Larger system sizes
are required to characterize these system-spanning smectic clusters, which we leave for future work.

\begin{figure}
\begin{center}
\includegraphics[width=0.45\textwidth]{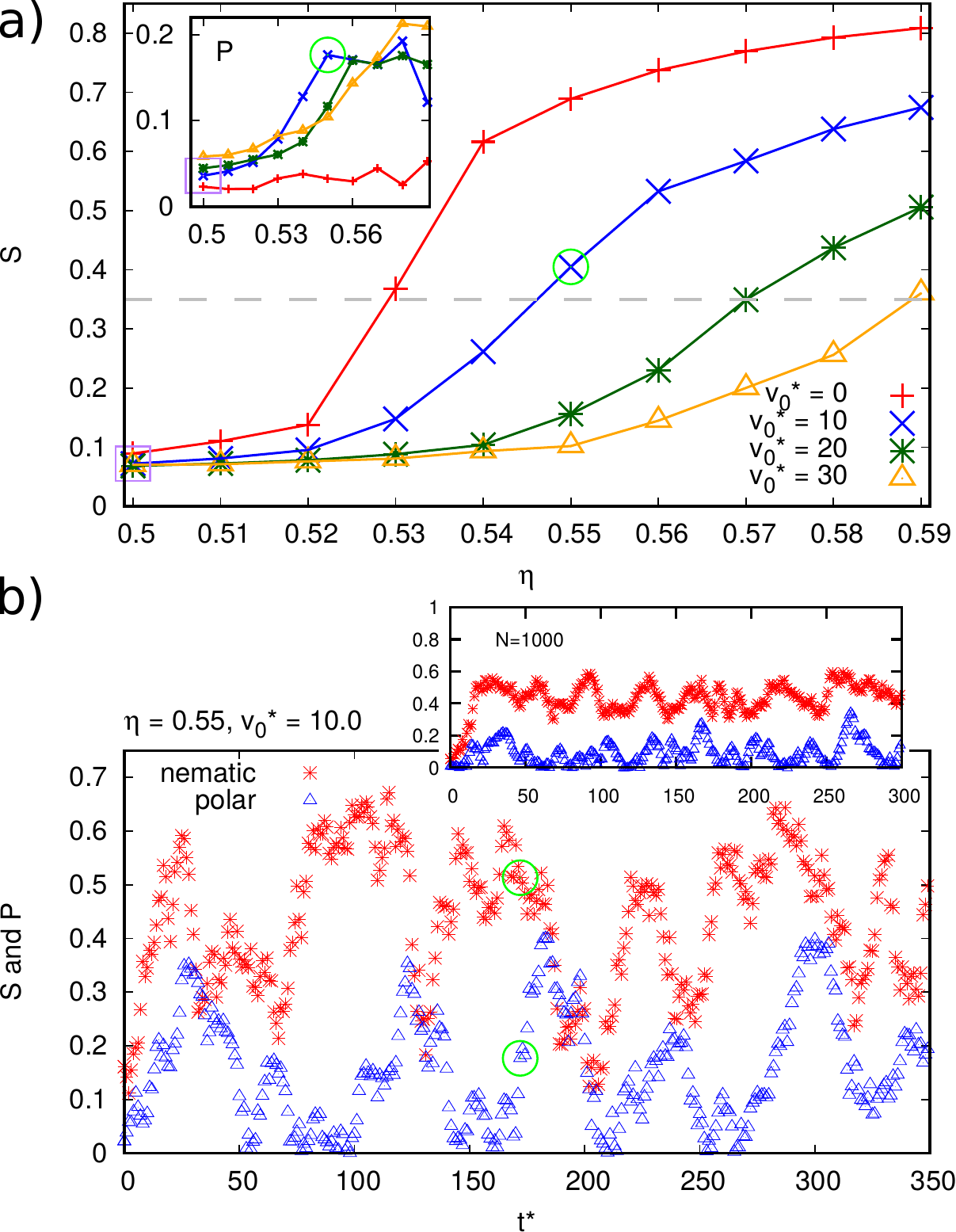}
\caption{Nematic $S$ and polar order parameter $P$ used to determine the phase behavior in the active rGB simulation.
The state points indicated by a square and a circle correspond to the snapshots in Figs.~\ref{fig:simulation_snapshots}a and b, respectively.
(a) Time average of $S$ (the inset shows $P$) for different velocities $v_0^*$ as a function of packing fraction $\eta$. 
The broken line indicates our threshold for determining the IN phase boundary.
Around this threshold, the error bars are comparable to the symbol size, whereas they become significantly smaller in the nematic phase
and are barely visible in the isotropic phase (see also the SI~\cite{supinfo_active_ellipsoids}).
(b) Time evolution $S(t)$ and $P(t)$ for $\eta=0.55$ and $v_0^*=10$.
 The inset shows the same plots for a larger system with $N=1000$ particles (see also the SI~\cite{supinfo_active_ellipsoids}).
}
\label{fig:order_parameter_small_activities}
\end{center}
\end{figure}

To understand the differences between the active nematic phase in the two models, let us first analyze the rGB system in some more detail.
In Fig.~\ref{fig:order_parameter_small_activities}a we show the
time-averaged global order parameters from Eq.~\eqref{eq:SP} at packing fractions $\eta$
close to the IN phase boundary in the active rGB system.
Following a path of state points by increasing $v_0$ at a fixed packing fraction 
leads to a decrease in the nematic order parameter $S$, which 
eventually falls below our chosen threshold $S_\text{t}$.  
Beyond this transition point, we classify the state as isotropic
and conclude that the activity destabilizes the nematic phase. 
The transition packing fraction shifts to higher values at higher activity.
We also observe a reduced slope of $S(\eta)$ at higher activity, 
which is why different threshold values $S_\text{t}$ would result in a slightly different phase boundary.
As illustrated by the behavior of $P$ 
in the inset of Fig.~\ref{fig:order_parameter_small_activities}a,
the emergence of nematic order is not associated with persistent global polar ordering.

To make a clear statement about the behavior of the active system, it is important to discuss the role of fluctuations. 
In the global isotropic phase it is well known \cite{Yang2010,Abkenar2013,Weitz2015} that there emerge local polar clusters with a critical size,
which increases upon increasing the activity or the density.
The local polar state depicted in Fig.~\ref{fig:phasediagram} for the rGB ellipsoids thus corresponds to a single cluster spanning the whole system.
On increasing the system size, the associated ``phase boundary'' shifts to higher $v_0$ for a given $\eta$, 
which consistently verifies that the polar state in our finite-size simulation
does not represent a true nonequilibrium phase with global order 
in an infinite system \cite{Yang2010,Abkenar2013,Weitz2015}. 
In the states which we characterize as nematic the polar fluctuations are much more prominent 
than one would expect for an isotropic phase with the same parameters.
 In fact, even in the actual isotropic phase found at the same density but \textit{higher} activity, the fluctuations are significantly weaker.
In Fig.~\ref{fig:order_parameter_small_activities}b we show the time evolution of
both order parameters associated with the nematic snapshot in Fig.~\ref{fig:simulation_snapshots}b. 
The pronounced temporal fluctuations near the transition result in slightly larger errors of the time-averaged values compared to the bulk phases
and also rationalize the decrease of the slope of $S(\eta)$ at higher activity, observed in Fig.~\ref{fig:order_parameter_small_activities}a.
Moving deeper into the nematic phase, the nematic order parameter can be determined quite accurately. 

In other words, we suspect that the fluctuations discussed above for the rGB ellipsoids are related to an enhancement 
of unphysical, finite-size induced self-interactions due to the persistent motion of the 
aligned rods in the nematic phase. 
However, the following considerations support our claim that the 
IN phase boundary depicted in Fig.~\ref{fig:HSC_GB_comparison} is generic.
Firstly, we stress that the observed long-time behavior is independent of the (either polar or isotropic) initial conditions.
Secondly, upon further increasing the activity, the nematic phase eventually turns into a distinct isotropic phase with significantly fewer fluctuations, 
which points to a well-defined phase transition even if the fluctuations in the nematic phase partially arise from finite-size effects.
Finally, both the lifetime and the magnitude of the described fluctuations decrease with increasing system size, as indicated in Fig.~\ref{fig:order_parameter_small_activities}b, 
whereas the average nematic order parameter is robust, i.e., the IN phase boundary in Fig.~\ref{fig:phasediagram} does not change.
We even found indications that the transition, i.e., the change of the nematic order parameter in Fig.~\ref{fig:order_parameter_small_activities}a, becomes sharper in a larger system.
 For more details on the finite-size effects and fluctuations in the rGB simulations see the SI~\cite{supinfo_active_ellipsoids}.

\begin{figure}
\centerline{\includegraphics[width=0.45\textwidth]{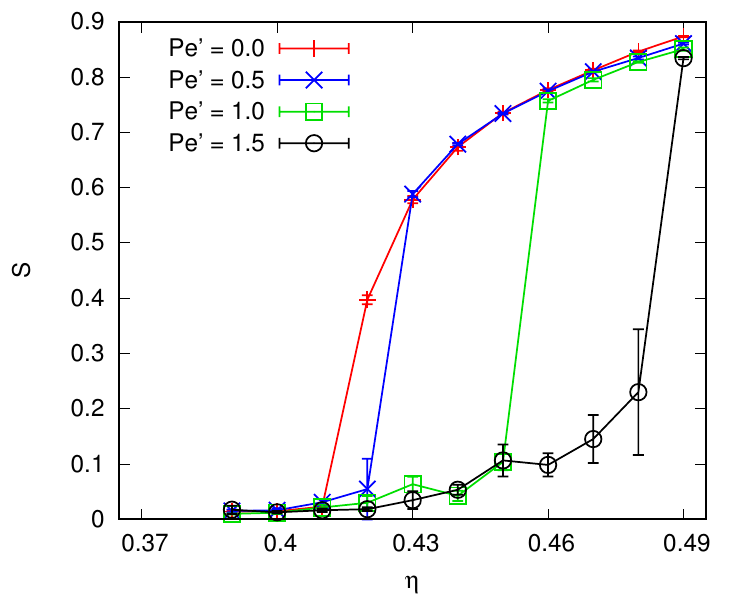}}
\caption{
	The nematic order parameter $S$ for HSC as a function of the packing fraction  $\eta$ for various swimming speeds $\mathrm{Pe}'$.
	The error bars indicate the standard deviation.
\label{fig:S2_of_eta_HSC}
}
\end{figure}

The above discussion is corroborated by our simulations of the HSC system, where we do not observe significant fluctuations of the global order parameters in the nematic phase,
which is similar to the equilibrium nematic order parameter, even relatively close to the phase boundary.
In particular, the IN transition is always rather sharp, as illustrated in Fig.~\ref{fig:S2_of_eta_HSC}. This is, at least partially, due to the much larger system
size of the HSC system. The error bars, which are a measure of the standard deviation (not the standard error), in the nematic order parameter increases drastically
in the isotropic phase at higher propulsion speed, especially $Pe'=1.5$ and near the transition. Large error bars are an indication of large fluctuations, such as those found in the rGB system.
However, for the HSC system, the fluctuations are much less pronounced than in the smaller rGB system.
 It is known that the IN transition is first order in equilibrium, although the coexistence region is very small~\cite{McGrother1996}. Since the region of bistability cannot completely disappear if an
infinitesimal propulsion speed is imposed, there must be a
(small) region of bistability at nonzero propulsion speed (at least for small $v_0^*$).
The jump in the order parameter, the magnitude of which is only weakly affected by the self-propulsion (see Fig.~\ref{fig:S2_of_eta_HSC}),
indicates that the IN transition remains discontinuous, but the expected region of bistability is smaller than our density resolution for all $v_0^*$
(including the equilibrium system).

Due to the different aspect ratios of the two types of particles and the resulting difference in friction,
the swimming speed at which the transition starts to shift towards higher densities is reduced.
In the isotropic phase near the IN transition, the nematic order parameter fluctuates strongly as a function of time for the larger swimming speeds.
This explains the raggedness of the curves
in Fig.~\ref{fig:S2_of_eta_HSC} for larger velocities (especially $\mathrm{Pe}'=1.5$).
When investigating by eye the snapshots of the isotropic phase in the HSC system, we made the following observations:
In the isotropic phase, as for the rGB model, polar clusters were found that increase in size
when increasing activity or density (however, we made sure that the polar clusters of HSC never span the system). 
In contrast, the typical nematic configurations of the HSC do not show strong local polar ordering,
even at swimming speeds where the isotropic phase clearly exhibits large polar clusters, compare Figs.~\ref{fig:simulation_snapshots}c and~\ref{fig:simulation_snapshots}d.

\section{Conclusions}
\label{sec:conclusion}

In conclusion, we identified in 3D and for small aspect ratios a homogeneous nematic phase, close to equilibrium, 
which can be clearly distinguished from the isotropic phase, even in a relatively small system. 
 By homogeneous, we mean that there are no appreciable inhomogeneities in the local density.
This nonequilibrium nematic phase is gradually destabilized by activity
and we observe no evidence for giant number fluctuations (but we cannot exclude the possibility entirely). 
Our finding is not sensitive to the precise particle shape or the details of the
interaction, provided the rods are short.
The activity-induced stabilization of the nematic phase predicted by mean-field
theory~\cite{Baskaran2008} for 2D is thus not universal.

 The reason for our new observations could be related to the shortness of the considered particles.
In systems of long rods, especially in 2D,
head-to-side collisions dominate and will rotate the particles 
towards either a parallel or anti-parallel orientation, 
 which can be used as an argument in favor of enhancing the nematic order rather than destroying it. 
As the aspect ratio is reduced, head-on collisions become increasingly frequent, 
generating disorder and destabilizing the nematic phase. 
Moreover, as the aspect ratio is reduced, the passive IN transition moves 
to higher densities, which further increases the  
relative importance of head-on collisions.
The consequences of dense clustering and correlations beyond the 
mean-field level have not been taken into account in previous theoretical studies~\cite{Baskaran2008}.
 The finding that the active nematic phase in 3D systems should be more stable than in 2D,
is still reasonable since in the latter case the rods have less directions in which they can escape upon a collision, so they would cluster more readily.

The investigation of the effect of low-level activity on established equilibrium states
should, of course, also be carried out with hydrodynamic interactions taken into account.
One fundamental question to be addressed is then whether there are major differences between 
 this more realistic model and our overdamped simulations. 
If momentum is conserved in such a more realistic model, is it possible that the nematic phase becomes unstable with respect to 
inhomogeneous flows with large wave length, as predicted by (linear) hydrodynamic theory~\cite{Simha2002,Marchetti2013}.
In our system, these large-wavelength instabilities are suppressed~\cite{Simha2002,Marchetti2013} by the friction and noise terms in
the equations of motion; as a result, momentum is not conserved. Similarly, walls in an experimental system also act to violate momentum conservation~\cite{Nishiguchi2017}.
It will be interesting to see to what extent these effects suffice to recover the behavior found in this work.
In any case, our work will provide an important benchmark to understand the role of the ignored hydrodynamic interactions in the near-equilibrium regime.
We thus hope that our work will motivate experiments on nematic phases of SPRs in near-equilibrium.
Such active liquid crystals could, for example, be constructed by rendering active a system of synthetic colloidal rods \cite{Kuijk2011}.
We are thus confident that the problem of active perturbations of equilibrium phases, 
 pioneered by our simulations (and similar efforts for other systems \cite{kapfer,Bialke2012}) 
is not only of pure theoretical interest. 

An open task for our overdamped simulations is to provide a
more fundamental quantitative understanding of the nature of the observed nonequilibrium IN transition
and the active nematic phase in particular.
Although the transition appears to remain of first order (as in equilibrium),
which we suspect from the sharp increase of the nematic order parameter in the HSC system,
a more careful analysis is required for a definite statement.
Since we have established a clear connection to the IN transition in equilibrium,
we do not believe that underlying mechanism driving this transition
is comparable to a liquid-vapor-like motility-induced phase separation \cite{Cates2015}.
The latter (not to be confused with polar clustering) could rather be observed within the isotropic region of the phase diagram, i.e.,
at lower density, higher activity and, perhaps, only for shorter rods.
In this sense, and to obtain clarity on the collision argument, it will also be of interest to study the influence of both the aspect ratio and interparticle interactions on the active IN phase boundary. 
To properly characterize the active nematic phase and the transition region
a detailed analysis of different pair correlation functions and the orientational distribution will be presented in future work.
Along these lines, we will also explore in detail the high-density region in the HSC system to conclusively argue about the existence of an active smectic phase.

The most important open task is, however, on the theoretical side.
It would be desirable to have a first-principles theoretical 
approach to confirm our surprising predictions of the activity dependence of the IN phase boundary, even if this is limited to 
low activities, close to equilibrium. 
One obvious possibility would be to develop a linear-response theory \cite{Sharma2016} for an anisotropic and active system.
While the phase behavior of spherical ABPs can be explained solely by effective attractions \cite{Cates2015,Schwarz-Linek2012,Wittmann2016} 
and that of active nematic rods by an effective (longer) aspect ratio \cite{Kraikivski2006},
an appropriate effective potential for SPRs should account for their characteristic broken up-down symmetry.
The most simplistic passive model system with this property consists of hard pear-shaped objects,
for which it has been detailed recently that the nematic phase
destabilizes with increasing deviation from ellipsoidal shape \cite{Schoenhoefer2017}.
This observation suggests an intuitive mapping 
to describe the IN transition in qualitative agreement with our simulations, which is yet to be quantified.
Another promising and possibly computationally efficient approach would be an implementation 
within dynamical density functional theory~\cite{Archer2004a} for anisotropic and active systems~\cite{Wittkowski2011},
which recently has been generalized also to microswimmers in a hydrodynamic medium \cite{Menzel2016}.

In conclusion, there is much opportunity for further experimental, theoretical and numerical studies of the active nematic phase of SPRs.  
Beyond the bulk system, these should also address the Frank elastic behavior, the response to (time-dependent) external fields and inhomogeneous 
systems in the presence of confining walls.

\section*{Acknowledgements}

The authors want to thank Sebastian Kapfer for a careful reading of the manuscript and valuable comments and
Hartmut L\"owen for helpful discussions.
R.~Wittmann and J.\ M.\ Brader acknowledge funding provided by the Swiss National Science Foundation,
M.\ Bott the support by the Swiss National Science Foundation through the National Center of Competence in Research Bio-Inspired Materials
 and F.\ Winterhalter acknowledges support by the Deutsche Forschungsgemeinschaft (DFG) through the Research Unit ``Geometry and Physics of Spatial Random Systems'' (GPSRS) under grant number ME1361/11.
Finally, the authors gratefully acknowledge the compute resources and support provided by the Erlangen Regional Computing Center (RRZE).


\appendix\newpage

\section{Numerical details}
\label{app:AppendixA}

In this appendix we give a detailed description of the Gay-Berne (GB) model, present how to calculate the forces and torque
and describe how to extract the required order parameters from the numerical data.

\subsection{The Gay-Berne model}

The Gay-Berne interaction potential for anisotropic particles is given by
\begin{align}
 \label{eq:gay_berne_potential_app}
 \phi_{\rm gb}(\mathbf{\hat{u}}_1, \mathbf{\hat{u}}_2, \mathbf{r}) &= \nonumber \\
 &4\epsilon(\mathbf{\hat{u}}_1, \mathbf{\hat{u}}_2, \mathbf{\hat{r}})
 \Biggl\{\biggl( \frac{\sigma_0}{r - \sigma(\mathbf{\hat{u}}_1, \mathbf{\hat{u}}_2, \mathbf{\hat{r}}) + \sigma_0}\biggr)^{12} \nonumber \\ 
 &- \biggl( \frac{\sigma_0}{r - \sigma(\mathbf{\hat{u}}_1, \mathbf{\hat{u}}_2, \mathbf{\hat{r}}) + \sigma_0}\biggr)^{6} \Biggr\}\,,
\end{align}
With the unit vectors $\mathbf{\hat{u}}_1$ and $\mathbf{\hat{u}}_2$ specifying the orientation of the interacting particles $1$ and $2$ and $\mathbf{r}$, $r$, $\mathbf{\hat{r}}$ their center to center-vector, -distance and -direction.
The attraction depth $\epsilon(\mathbf{\hat{u}}_1, \mathbf{\hat{u}}_2, \mathbf{\hat{r}})$ and the range $\sigma(\mathbf{\hat{u}}_1, \mathbf{\hat{u}}_2, \mathbf{\hat{r}})$ of the particle interaction are dependent on the orientation. 

The shape of Gay-Berne particles is defined through an anisotropy parameter
\begin{equation*}
 \chi = \frac{1/\kappa^2-1}{1/\kappa^2+1}, \qquad \kappa = \sigma_s/\sigma_e.
\end{equation*}
Here $\sigma_e$ is the ``length'' of the particle defined by the end-to-end interaction and $\sigma_s$ the ``width'' of the particle defined though the side-to-side interaction.
For infinitely long cigar shaped particles $(\sigma_e \to \infty)$ the anisotropy parameter $\chi \to 1$, in contrast for infinitely thin oblate-like particles $(\sigma_s \to 0)$ we have $\chi \to -1$.

The orientation dependent interaction range is given by
\begin{align*}
\sigma(\mathbf{\hat{u}}_1, \mathbf{\hat{u}}_2, \mathbf{\hat{r}}) &= \sigma_0 \Biggl[ 1 - \frac{1}{2}\chi \biggl \{ \frac{(\mathbf{\hat{r}} \cdot \mathbf{\hat{u}}_1 + \mathbf{\hat{r}} \cdot \mathbf{\hat{u}}_2)^2}{1 + \chi(\mathbf{\hat{u}}_1 \cdot \mathbf{\hat{u}}_2)} \\
&+ \frac{(\mathbf{\hat{r}} \cdot \mathbf{\hat{u}}_1 - \mathbf{\hat{r}} \cdot \mathbf{\hat{u}}_2)^2}{1 - \chi(\mathbf{\hat{u}}_1 \cdot \mathbf{\hat{u}}_2)} \biggr \} \Biggr]^{-\frac{1}{2}}.
\end{align*}
The well depth is defined as
\begin{equation*}
 \epsilon(\mathbf{\hat{u}}_1, \mathbf{\hat{u}}_2, \mathbf{\hat{r}}) = \epsilon_0 \, \epsilon(\mathbf{\hat{u}}_1, \mathbf{\hat{u}}_2)^\nu \, \epsilon'(\mathbf{\hat{u}}_1, \mathbf{\hat{u}}_2, \mathbf{\hat{r}})^{\mu},
\end{equation*}
\begin{equation*}
\epsilon(\mathbf{\hat{u}}_1, \mathbf{\hat{u}}_2) = \bigl\{ 1 - \chi^2  (\mathbf{\hat{u}}_1, \mathbf{\hat{u}}_2)^2 \bigr\}^{-1/2},
\end{equation*}
\begin{align*}
\epsilon'(\mathbf{\hat{u}}_1, \mathbf{\hat{u}}_2, \mathbf{\hat{r}}) &= 1 - \frac{\chi'}{2} \biggl \{ \frac{(\mathbf{\hat{r}} \cdot \mathbf{\hat{u}}_1 + \mathbf{\hat{r}} \cdot \mathbf{\hat{u}}_2)^2}{1 + \chi'(\mathbf{\hat{u}}_1 \cdot \mathbf{\hat{u}}_2)} \\
&+ \frac{(\mathbf{\hat{r}} \cdot \mathbf{\hat{u}}_1 - \mathbf{\hat{r}} \cdot \mathbf{\hat{u}}_2)^2}{1 - \chi'(\mathbf{\hat{u}}_1 \cdot \mathbf{\hat{u}}_2)} \biggr \},
\end{align*}
with the parameter $\chi'$ describing the anisotropy in the well depth:
\begin{equation*}
\chi' = \frac{1 - \kappa'^{\frac{1}{\mu}}}{1 + \kappa'^{\frac{1}{\mu}}}, \qquad \kappa' = \frac{\epsilon_s}{\epsilon_e}.
\end{equation*}
In the limit of spherical particles i.e. $\sigma_e = \sigma_s$, $\epsilon_e = \epsilon_s$ one finds $\chi = 0$ and $\chi' = 0$ and therefore the Gay-Berne potential becomes a regular Lennard-Jones interaction.

In our study we employed a soft repulsive WCA-like version of the Gay-Berne interaction, which is obtained by shifting and truncating the Gay-Berne potential:
\begin{align}\label{eq:wca_gay_berne_potential_app}
 \phi_{\rm wca}&(\mathbf{\hat{u}}_1,\mathbf{\hat{u}}_2,\mathbf{r}) \nonumber \\
 &=\begin{cases}
    \phi_{\rm gb}(\mathbf{\hat{u}}_1,\mathbf{\hat{u}}_2,\mathbf{r}) + \epsilon(\mathbf{\hat{u}}_1,\mathbf{\hat{u}}_2,\mathbf{r}) \;\; r\!<\!r_{\rm min}\\
    0 \qquad \qquad \qquad \qquad \qquad \;\;\;\;\;\;\;r \ge r_{\rm min},
    \end{cases}
\end{align}
where $r_{\rm min}(\mathbf{\hat{u}}_1,\mathbf{\hat{u}}_2,\mathbf{\hat{r}}) = 2^{1/6}\sigma_0 
+ \sigma(\mathbf{\hat{u}}_1,\mathbf{\hat{u}}_2,\mathbf{\hat{r}}) - \sigma_0$ is the 
minimum of the Gay-Berne interaction.

\subsection{Calculation of Force and Torque}

The expression to calculate the force and torque we denote for the regular Gay-Berne interaction (\ref{eq:gay_berne_potential_app}). 
For this we introduce the scaled variable
\begin{equation*}
 R = \frac{r - \sigma(\mathbf{\hat{u}}_1, \mathbf{\hat{u}}_2, \mathbf{r}) + \sigma_0 }{\sigma_0}
\end{equation*}
and the interaction potential can be written as
\begin{equation*}
 \phi_{\rm gb}(\mathbf{\hat{u}}_1, \mathbf{\hat{u}}_2, \mathbf{r}) = 4\epsilon(\mathbf{\hat{u}}_1, \mathbf{\hat{u}}_2, \mathbf{\hat{r}}) \Biggl\{\biggl(\frac{1}{R}\biggr)^{12} - \biggl(\frac{1}{R}\biggr)^{6} \Biggr\}.
\end{equation*}
Furthermore we define a function $g(X)$ for the relative orientation of two interacting particles
\begin{equation*}
 g(X) = 1 - \frac{X}{2} \biggl \{ \frac{(\mathbf{\hat{r}} \cdot \mathbf{\hat{u}}_1 + \mathbf{\hat{r}} \cdot \mathbf{\hat{u}}_2)^2}{1 + X (\mathbf{\hat{u}}_1 \cdot \mathbf{\hat{u}}_2)} + \frac{(\mathbf{\hat{r}} \cdot \mathbf{\hat{u}}_1 - \mathbf{\hat{r}} \cdot \mathbf{\hat{u}}_2)^2}{1 - X (\mathbf{\hat{u}}_1 \cdot \mathbf{\hat{u}}_2)} \biggr \}.
\end{equation*}
Hence we have
\begin{equation*}
 \sigma(\mathbf{\hat{u}}_1, \mathbf{\hat{u}}_2, \mathbf{r}) = \sigma_0 \, g(\chi)^{-1/2}
\end{equation*}
and
\begin{equation*}
 \epsilon'(\mathbf{\hat{u}}_1, \mathbf{\hat{u}}_2, \mathbf{\hat{r}})  = g(\chi').
\end{equation*}
When changing the distance vector $\mathbf{r}$ between two particles the interparticle vector $\mathbf{\hat{r}}$ changes as well, which can be made explicit in $g(X)$
\begin{equation*}
 g(X) = 1 - \frac{X}{2r^2} \biggl \{ \frac{(\mathbf{r} \cdot \mathbf{\hat{u}}_1 + \mathbf{r} \cdot \mathbf{\hat{u}}_2)^2}{1 + X (\mathbf{\hat{u}}_1 \cdot \mathbf{\hat{u}}_2)} + \frac{(\mathbf{r} \cdot \mathbf{\hat{u}}_1 - \mathbf{r} \cdot \mathbf{\hat{u}}_2)^2}{1 - X (\mathbf{\hat{u}}_1 \cdot \mathbf{\hat{u}}_2)} \biggr \}.
\end{equation*}

\subsubsection*{Force}

The force is given by $\mathbf{F} = -\nabla \phi_{\rm gb}$. We here denote the expression for the force in $x$-direction:
\begin{align}
\label{eq:gayberneforce}
 -F_x  = \frac{\partial \phi_{\rm gb}}{\partial x} &= 4 \frac{\partial \epsilon}{\partial x} \Biggl\{\biggl(\frac{1}{R}\biggr)^{12} - \biggl(\frac{1}{R}\biggr)^{6} \Biggr\} \\
 &+ 4\epsilon \Biggl\{\biggl(\frac{6}{R}\biggr)^{7} - \biggl(\frac{12}{R}\biggr)^{13} \Biggr\}\frac{\partial R}{\partial x},
\end{align}
with
\begin{equation*}
 \frac{\partial \epsilon}{\partial x} = \epsilon_0 \epsilon(\mathbf{\hat{u}}_1, \mathbf{\hat{u}}_2)^{\nu} \mu \epsilon'(\mathbf{\hat{u}}_1, \mathbf{\hat{u}}_2, \mathbf{\hat{r}})^{\mu-1} \frac{\partial g(\chi')}{\partial x},
\end{equation*}
\begin{equation*}
\frac{\partial R}{\partial x} = \frac{1}{\sigma_0} \biggl(\frac{x}{r} - \frac{\partial \sigma}{\partial x} \biggr),
\end{equation*}
and
\begin{equation*}
 \frac{\partial \sigma}{\partial x} = -\frac{1}{2} \sigma_0 g(\chi)^{-3/2} \frac{\partial g(\chi)}{\partial x}.
\end{equation*}
Finally the derivative of the orientation function $g(X)$ is given by
\begin{align*}
 \frac{\partial g(X)}{\partial x} &= \frac{x X}{r^4} \biggl \{ \frac{(\mathbf{r} \cdot \mathbf{\hat{u}}_1 + \mathbf{r} \cdot \mathbf{\hat{u}}_2)^2}{1 + X (\mathbf{\hat{u}}_1 \cdot \mathbf{\hat{u}}_2)} + \frac{(\mathbf{r} \cdot \mathbf{\hat{u}}_1 - \mathbf{r} \cdot \mathbf{\hat{u}}_2)^2}{1 - X (\mathbf{\hat{u}}_1 \cdot \mathbf{\hat{u}}_2)} \biggr \} \\
 &- \frac{X}{r^2} \biggl \{ \frac{(\mathbf{r} \cdot \mathbf{\hat{u}}_1 + \mathbf{r} \cdot \mathbf{\hat{u}}_2)}{1 + X (\mathbf{\hat{u}}_1 \cdot \mathbf{\hat{u}}_2)} (\hat{u}_1^x + \hat{u}_2^x) \\
 &+ \frac{(\mathbf{r} \cdot \mathbf{\hat{u}}_1 - \mathbf{r} \cdot \mathbf{\hat{u}}_2)}{1 - X (\mathbf{\hat{u}}_1 \cdot \mathbf{\hat{u}}_2)} (\hat{u}_1^x - \hat{u}_2^x) \biggr \}.
\end{align*}
The force in y- and z-direction can be calculated equivalently.

\subsubsection*{Torque}

Due to the angular dependence of the Gay-Berne potential particles experience torque.
So far we only determined the center to center force.
We can calculate the torque from an equivalent force $\mathbf{E}$ acting on a point at unit distance from the center of the particles.
This equivalent force can be calculated from the derivative of the potential with respect to the unit vector
\begin{equation*}
 \mathbf{E} = - \left (
		\begin{array}{c}
                 \partial \phi_{\rm gb}/ \partial \hat{u}_1^x\\
                 \partial \phi_{\rm gb}/ \partial \hat{u}_1^y\\
                 \partial \phi_{\rm gb}/ \partial \hat{u}_1^z
                \end{array}
                \right )
\end{equation*}
Again we denote the derivatives of $\phi_{\rm gb}$ with respect to $\hat{u}_1^x$, but in other directions and for particle $2$ one obtains equivalent results.
\begin{align*}
 -E_x = \frac{\partial \phi_{\rm gb}}{\partial \hat{u}_1^x} &= 4 \frac{\partial \epsilon}{\partial \hat{u}_1^x} \Biggl\{\biggl(\frac{1}{R}\biggr)^{12} - \biggl(\frac{1}{R}\biggr)^{6} \Biggr\}\\
 &+ 4 \epsilon \Biggl\{\frac{6}{R^{7}} - \frac{12}{R^{13}} \Biggr\} \frac{\partial R}{\partial \hat{u}_1^x}.
\end{align*}
Where
\begin{eqnarray*}
\frac{\partial \epsilon(\mathbf{\hat{u}}_1, \mathbf{\hat{u}}_2, \mathbf{\hat{r}})}{\partial \hat{u}_1^x} &=& \epsilon_0 \nu \epsilon(\mathbf{\hat{u}}_1, \mathbf{\hat{u}}_2)^{\nu-1}\frac{\partial \epsilon(\mathbf{\hat{u}}_1, \mathbf{\hat{u}}_2)}{\partial \hat{u}_1^x}\epsilon'(\mathbf{\hat{u}}_1, \mathbf{\hat{u}}_2, \mathbf{\hat{r}})^{\mu} \\
 &+& \epsilon_0 \nu \epsilon(\mathbf{\hat{u}}_1, \mathbf{\hat{u}}_2)^{\nu} \mu \epsilon'(\mathbf{\hat{u}}_1, \mathbf{\hat{u}}_2, \mathbf{\hat{r}})^{\mu-1} \frac{\partial g(\chi')}{\partial \hat{u}_1^x}, 
\end{eqnarray*}
\begin{equation*}
 \frac{\partial \epsilon(\mathbf{\hat{u}}_1, \mathbf{\hat{u}}_2)}{\partial \hat{u}_1^x} = \epsilon(\mathbf{\hat{u}}_1, \mathbf{\hat{u}}_2)^3 \chi^2 \mathbf{\hat{u}}_1 \cdot\mathbf{\hat{u}}_2 \hat{u}_2^x,
\end{equation*}
\begin{equation*}
 \frac{\partial R}{\partial \hat{u}_1^x} = \frac{1}{2} \biggl( \frac{\sigma(\mathbf{\hat{u}}_1, \mathbf{\hat{u}}_2, \mathbf{r})}{\sigma_0} \biggr)^3 \frac{\partial g(\chi)}{\partial \hat{u}_1^x}.
\end{equation*}
And the derivative of the orientation function is given by
\begin{align*}
&\frac{\partial g(X)}{\partial \hat{u}_1^x} = \\
&-\frac{X}{2} \Biggl [ \hat{r}_x \biggl \{ \frac{2(\mathbf{\hat{r}} \cdot \mathbf{\hat{u}}_1 + \mathbf{\hat{r}} \cdot \mathbf{\hat{u}}_2)}{1 + X (\mathbf{\hat{u}}_1 \cdot \mathbf{\hat{u}}_2)} + \frac{2(\mathbf{\hat{r}} \cdot \mathbf{\hat{u}}_1 - \mathbf{\hat{r}} \cdot \mathbf{\hat{u}}_2)}{1 - X (\mathbf{\hat{u}}_1 \cdot \mathbf{\hat{u}}_2)} \biggr \} \\ 
&+ X \hat{u}_2^x \biggl \{ \frac{(\mathbf{\hat{r}} \cdot \mathbf{\hat{u}}_1 - \mathbf{\hat{r}} \cdot \mathbf{\hat{u}}_2)^2}{(1 - X (\mathbf{\hat{u}}_1 \cdot \mathbf{\hat{u}}_2))^2} - \frac{(\mathbf{\hat{r}} \cdot \mathbf{\hat{u}}_1 + \mathbf{\hat{r}} \cdot \mathbf{\hat{u}}_2)^2}{(1 + X (\mathbf{\hat{u}}_1 \cdot \mathbf{\hat{u}}_2))^2} \biggr \} \Biggr].
 \end{align*}
The second term in the first bracket changes sign when taking the derivative with respect to orientation of particle 2.
Finally we obtain the torque by the cross product of $\mathbf{E}$ and the orientation vector $\mathbf{\hat{u}}_1$,
\begin{equation}
\label{eq:gaybernetorque}
\boldsymbol{T} = \mathbf{\hat{u}}_1 \times \mathbf{E}. 
\end{equation}

\subsection{Calculation of order parameters} \label{sec:app_order_pars}

In this appendix we provide information on how to extract the order parameters from the simulation data.
The orientational behavior of an ensemble of $N$ anisotropic particles can be analyzed using an order parameter $S$, which is defined as
\begin{equation*}
 S = \sum_{i=1}^N \frac{3 \cos^2\beta_i - 1}{2N},
\end{equation*}
where $\beta_i$ is the angle between the orientation-vector of particle $i$ and the nematic director (unit vector indicating the mean orientation of the particles).
The order parameter can take values between $0$ and $1$, where $S=0$ indicates that the system is in a fully isotropic state with random orientation and $S=1$ means perfect alignment of the particles.
However in simulations the nematic director is not known a priori. Following reference \cite{Eppenga1984} we consider a tensorial order parameter 
\begin{equation*}
 Q_{\alpha \beta} =  \frac{1}{N} \sum_{i=1}^N \frac{3}{2} \hat{u}_{i\alpha}\hat{u}_{i\beta} - \frac{1}{2}\delta_{\alpha\beta}, \qquad \alpha,\beta = x,y,z.
\end{equation*}
This 2nd-rank tensor can immediately be computed from the single particle orientations $\mathbf{\hat{u}}$. 
It has three eigenvalues, of which the largest is the order parameter and the corresponding eigenvector the nematic director $\mathbf{\hat{n}}$.

In systems with polar order, the nematic order parameter is also nonzero and the nematic director lies along the direction of polar order.
Thus, we can reuse the nematic director to calculate the polar order parameter, when we define it as
\begin{equation}
P = \frac{1}{N}\Big| \sum_{i=1}^N \mathbf{\hat{u}}_i \cdot \mathbf{\hat{n}} \,\Big|.
\end{equation}
Again the polar order parameter ranges between 0 and 1, indicating no polar alignment and perfect polar ordering, respectively.

\bibliography{PhD_bibliography.bib}

\end{document}